\documentclass[a4paper,fleqn,usenatbib]{mnras}

\usepackage[T1]{fontenc}
\usepackage{ae,aecompl}

\usepackage{graphicx}	
\usepackage{amsmath}	
\usepackage{amssymb}	
\usepackage{bm}
\usepackage{color}
\usepackage{ulem}
\usepackage{cancel}

\title[Convection and pulsation stability of stars]{Turbulent convection and pulsation stability of stars --
III. Non-adiabatic oscillations of red giants}

\author[D. R. Xiong et al.]{
D. R. Xiong,$^{1}$
L. Deng,$^{2}$
and C. Zhang$^{2}$
\\
$^{1}$Purple Mountain Observatory, Chinese Academy of Sciences, Nanjing 210008, China\\
$^{2}$Key Laboratory of Optical Astronomy, National Astronomical Observatories, Chinese Academy of Sciences, Beijing, 100101, China
}

\date{Accepted 2018 July 16. Received 2018 July 12; in original form 2017 July 24}

\pubyear{2018}

\begin{document}
\label{firstpage}
\pagerange{\pageref{firstpage}--\pageref{lastpage}}
\maketitle

\begin{abstract}
We have computed linear non-adiabatic oscillations of luminous red giants using a non-local and anisotropic time-dependent theory of convection. The results show that low-order radial modes can be self-excited. Their excitation is the result of radiation and the coupling between convection and oscillations. Turbulent pressure has important effects on the excitation of oscillations in red variables.
\end{abstract}




\section{Introduction}\label{sec1}

In the H-R diagram, there are a lot of pulsating red variables in the low-temperature area to the right of the Cepheid instability strip. They have the largest number among all known types of pulsating variables, and yet we know little about them. In the General Catalogue of Variable Stars \citep{GCVS}, luminous red variables, usually known as long-period variables (LPVs), are divided into three types according to the regularity of their light curves: Miras, semi-regular variables (SRVs) and irregular variables. The study of red variables has made considerable progress in the past two decades with the help of photometric observations from projects like MACHO, OGLE and 2MASS. \citet{Wood1999} and \citet{Wood2000} found 5 ridges (sequences A--E) in the period-luminosity (PL) diagram of luminous red variables in the Large Magellanic Cloud (LMC). OGLE observations showed more complex structures in the PL diagram, and at least 14 ridges could be defined \citep{KB2003, WEP2004, SUK2004, SUK2005, SDU2007}. \citet{SUK2004} found that stars with the primary periods on sequence A rarely had their second and third dominant periods on sequences C and C$'$, and stars with the primary periods on sequence C rarely had their second and third dominant periods on sequence A. This indicates that stars with their primary periods on sequence A are a special type of red variables. They are named OGLE Small Amplitude Red Giants (OSARGs), but their origin is still unclear.

At first \citet{Wood1999} thought that variable luminous red giants in MACHO observations were asymptotic-giant-branch (AGB) stars, but later \citet{KB2003} and \citet{ITM2004} found evidence showing that there were luminous red variables on both the AGB and the red-giant branch (RGB). Their PL sequences were parallel with slight offsets.

\citet*{TSI2013} compared observed period ratios of luminous red giants with theoretical ones and concluded that OSARGs were radial and non-radial p modes. \citet{Wood2015} reached a similar conclusion, and identified the stars on sequences C and C$'$ as predominantly radial pulsators; the radial pulsation modes associated with A$'$, A, B, C$'$ and C were the fourth, third, second and first overtones and the fundamental modes, respectively. However, \citet{Trabucchi2017} found that both sequences B and C$'$ corresponded to first-overtone pulsation.

For a long time, convection has been considered to be a pure damping mechanism of stellar oscillations. The excitation mechanism of pulsating red variables has been a long-standing theoretical problem. There are two different opinions regarding the excitation mechanism of OSARGs. The prevailing opinion is that they are stochastically excited, just like solar-like oscillations \citep{CDKM2001, SDU2007, TSI2013}. However, \citet{XD2013} came to the conclusion that there were no essential differences between the excitation mechanisms of oscillations in OSARGs and Miras. They were both the results of radiation and convective coupling. In this paper, we aim at probing the excitation mechanism for luminous red giants observed by MACHO and OGLE. In section \ref{sec2}, we briefly introduce theoretical stellar models and the scheme of computation. The results of computation of linear non-adiabatic oscillations are given in section 3, and the excitation mechanism of oscillations is discussed in section 4. We summarize our results in section 5.

\section{Theoretical models and scheme of computation}\label{sec2}

It is well known that traditionally there are 4 equations for stellar structure: the conservation of mass, momentum and energy, and the equation of radiative transfer. Convection is treated with the mixing-length theory \citep[MLT;][]{MLT}. MLT is a phenomenological theory based on the analogy between turbulent convection and kinetic theory of gas molecules. In fact turbulence is much more complex than motions of gas molecules. Most of the time, gas molecules are free except when they collide. Their mean free path is far less than the characteristic length of the average fluid field change. Therefore, the molecule transport process (molecular viscosity, molecular heat conduction, molecular diffusivity, \textbf{etc.}) can be treated with a local theory. On the contrary, turbulent elements are in continuous interaction, and the characteristic length of turbulent elements and the average field change are comparable. Therefore, the convective transport process is a non-local phenomenon. Nonlinearity and non-locality of fluid motions are two main difficulties in the study of convection. The fundamental shortcoming of the local MLT is that it is not a dynamic theory following the hydrodynamic equations, therefore it cannot correctly describe the dynamic behaviours of turbulent convection. This shortcoming is very serious, or even intolerable, in dealing with dynamic problems of non-local and time-dependent convection. To solve this problem, we have developed a non-local and anisotropic time-dependent theory of convection based on hydrodynamic equations and turbulence theory \citep{X1989, XCD1997, DXC2006}. There are 8 partial differential equations in our complete set of equations for chemically homogeneous stellar structure and oscillations. 4 of them are the equations of mass conversation, momentum conversation, energy conversation, and radiative transfer. When convection sets in, the second-order auto-correlation of turbulent velocity (the Reynolds stress) emerges in the momentum conversation equation, and the second-order cross-correlation of turbulent velocity and temperature (the convective enthalpy flux) appears in the energy conversation equation. The other 4 equations are the dynamic equations of the auto- and cross-correlations of turbulent velocity and temperature. There are 3 convective parameters ($c_1$, $c_2$, $c_3$), which are connected to dissipation, diffusion and anisotropy of turbulent convection, respectively. These parameters can be calibrated by using the structure of the solar convection zone, the turbulent velocity and temperature fields of the solar atmosphere, the solar lithium abundance and numerical simulations of hydrodynamics. From a pure theoretical point of view, these parameters cannot be a group of constants. They depend not only on stellar luminosity and effective temperature, but also on depth (radius) inside stars. Fortunately, our study shows that adopting a group of parameters ($c_1$, $c_2$, $c_3$)=(0.64, 0.32, 3.0) as calibrated using the Sun, we can reproduce almost all of the instability strips of pulsating stars in the H-R diagram. Moreover, within a wide range of these parameters stellar pulsation stability is not sensitive to the change of parameters, and is almost independent of the specific values of ($c_1$, $c_2$, $c_3$) \citep[][; here after Paper I]{tcpss1}.

We have calculated both the radial and nonradial oscillations of luminous red giants. In the present paper we discuss only the radial oscillations. The nonradial oscillations will be studied in detail in future work. For stellar radial oscillations, the effects of the core area can be neglected except for its gravity field. Therefore, envelope models are enough for our analysis. The number of equations for static structures and linear radial non-adiabatic oscillations in term of completely non-local and anisotropic theory is 12, while the number of equations for linear nonradial non-adiabatic oscillations is 16. In Paper I we gave the complete set of equations for stellar structure and oscillations as well as a brief description of our non-local and anisotropic time-dependent theory of convection. For detailed derivations of the equations please refer to our previous work \citep{X1989, XCD1997, DXC2006}.

Computing envelope models in the completely non-local and anisotropic convection theory is very difficult and time-consuming. We use quasi-anisotropic non-local convection models as equilibrium models for pulsation calculations. Our study shows that not only quasi-anisotropic non-local convection models approach the completely anisotropic ones very well in terms of the $T-P$ structure and turbulent velocity-temperature fields, but also their results of pulsation stability are consistent (see Paper I).

\section{Results of theoretical computation}\label{sec3}

\begin{figure}
\includegraphics[width=\columnwidth]{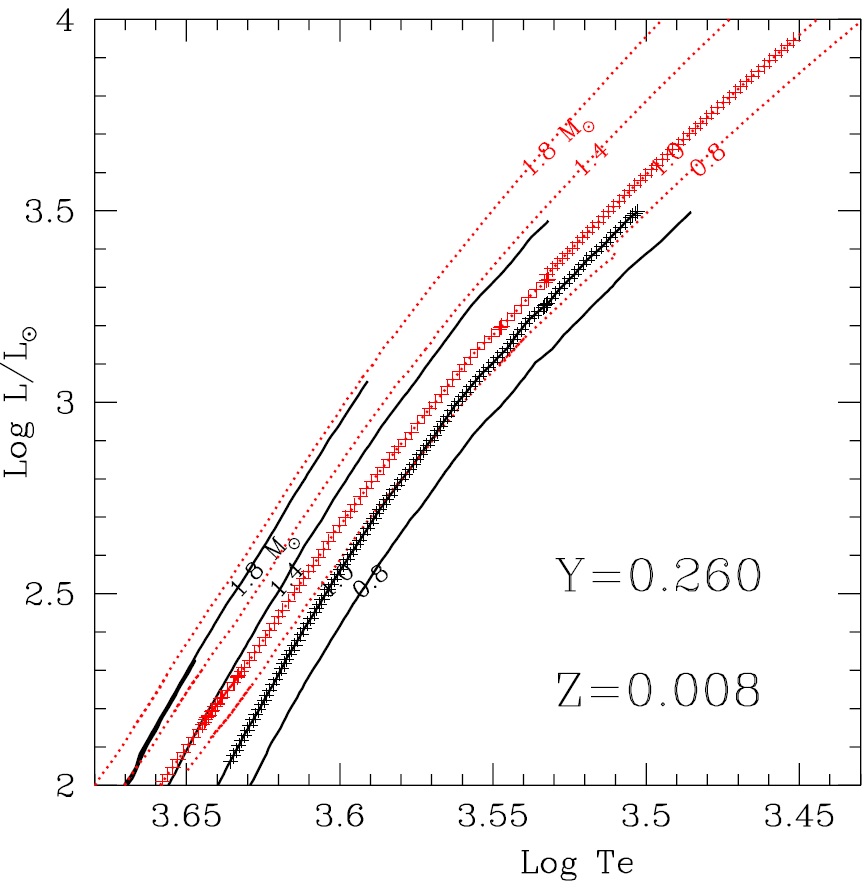}
\caption{RGB (black solid lines) and AGB (red dotted lines) evolutionary tracks with $M=0.8$, $1.0$, $1.4$ and $1.8\,\mathrm{M}_\odot$ in the H-R diagram. Pluses mark the positions of the $M=1.0\,\mathrm{M}_\odot$ equilibrium models in Figure~\ref{fig2}.}
\label{fig1}
\end{figure}

Adopting the scheme described in last section, we computed the radial and non-radial linear non-adiabatic oscillations of RGB and AGB models. The evolutionary models were from the Padova code \citep{Padova2008, Padova2009}, as shown in Figure~\ref{fig1}. The initial chemical abundances of LMC were $Y$=0.26, $Z$=0.008. Considering the lack of models in the thermal pulsating AGB (TP-AGB) phase and the uncertainty of mass loss, AGB models only included those from the horizontal branch to the beginning of the TP-AGB. The high-luminosity AGB stars with $L/\mathrm{L}_\odot \ga 3.3$ were extrapolated into the TP-AGB phase without mass loss. Therefore the results of oscillation computations of high-luminosity AGB stars are not very reliable, and can only be used as reference. We computed 150 non-local convective envelope models along each evolutionary track, then calculated their linear non-adiabatic oscillations for luminous RGB and AGB models of 0.8--1.8\,$\mathrm{M}_\odot$. The surface boundary of static envelop models and oscillation computations was at $\tau=10^{-3}$, where $\tau$ is the optical depth. The bottom boundary was set deep enough: the bottom temperature $T_\mathrm{b} \sim 8\times 10^6\,\mathrm{K}$ or fractional radius $r_\mathrm{b}/R_0 \sim 0.01$. A slightly modified version of the Mihalas-Hummer-D\"appen (MHD) equation of state \citep{MHD1,MHD2,MHD3}, and OPAL tabular opacity \citep{OPAL92} complemented by low-temperature tabular opacity \citep{LTO94} were adopted. All the computations of static models and oscillations followed the Henyey's algorithm \citep*{HFG1964}.

\begin{figure}
\includegraphics[width=\columnwidth]{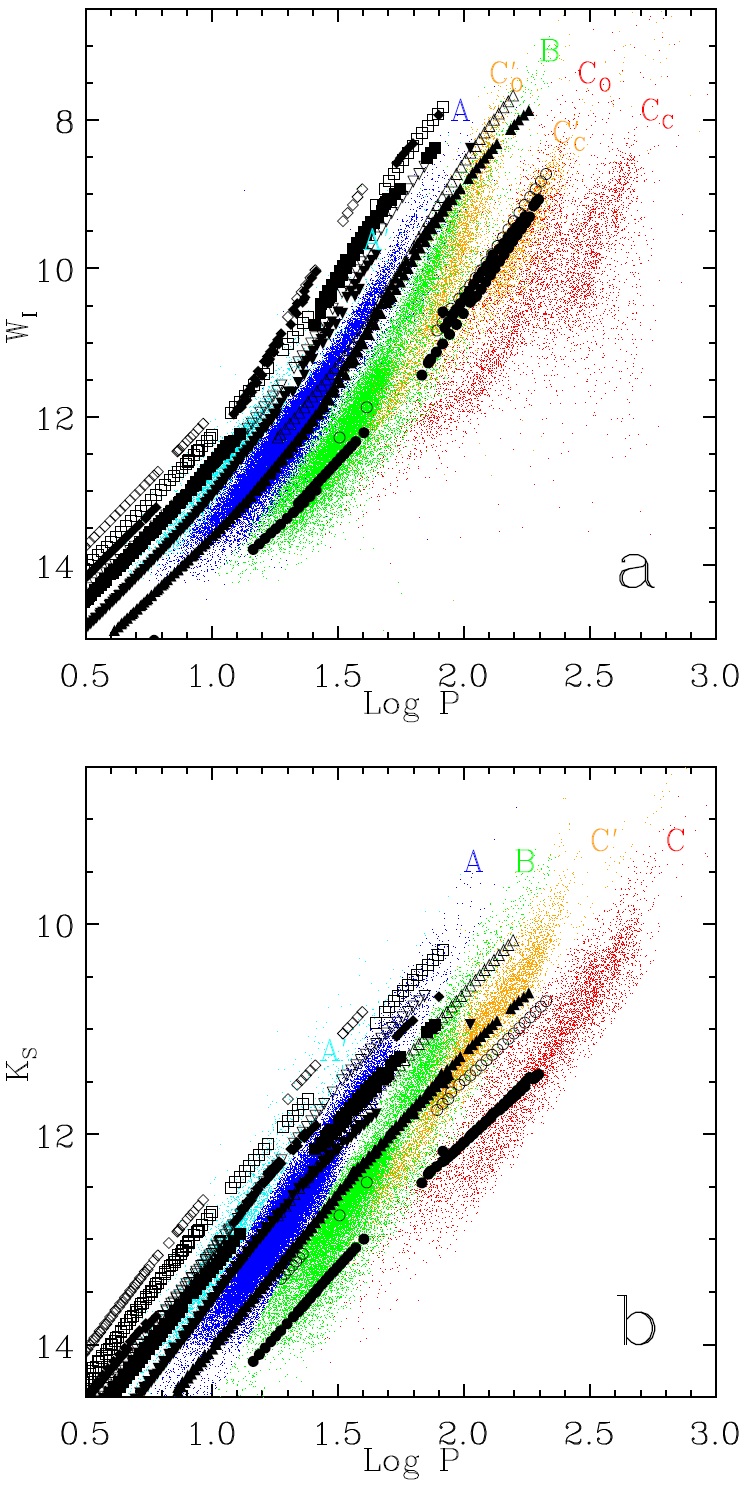}
\caption{Sequences of LMC LPVs in the $W_I-\log P$ plane (panel a) and $K_\mathrm{S}-\log P$ plane (panel b). Stars on sequences A$'$, A, B, C$'$, and C are shown as cyan, blue, green, orange, and red points, respectively. Sequence D is not shown. The filled and open black symbols show the theoretical PL relations for $1.0\,\mathrm{M}_\odot$ and $1.8\,\mathrm{M}_\odot$ evolutionary models, respectively. Circles, triangles, inverse triangles, squares, and diamonds are the pulsationally unstable radial fundamental modes, the first, second, third, and fourth overtones, respectively. }
\label{fig2}
\end{figure}

In order to compare with observations, stellar parameters were converted to $UBVRIJHK$ magnitudes using the transformation program provided by \citet{WL2011}. The $JHK$ colors were then converted to the 2MASS system using transformations given by \citet{Carpenter2001}. The adopted LMC distance module was 18.5 mag \citep{PGG2013}.

The observed sequences of luminous red variables in the LMC are shown in Figure \ref{fig2} using data from the OGLE-III catalogue of LPVs in the LMC \citep{SUS2009}. The abscissas are the logarithmic period (in days) $\log P$. The ordinate in panel a is the reddening-free Wesenheit index $W_I$, defined as
\begin{equation}
W_I=I-1.55(V-I).
\label{eq1}
\end{equation}
The ordinate in panel b is the 2MASS magnitude $K_\mathrm{S}$. The catalogue contains 79200 OSARGs, 11128 SRVs and 1667 Miras. In Figure \ref{fig2} only the primary period of each star is plotted. Stars on sequences A$'$, A, B, C$'$, and C are shown as cyan, blue, green, orange and red points, respectively. The subscripts O and C in panel a are used to distinguish between oxygen-rich and carbon-rich SRVs and Miras.

The theoretical PL relations of low-order radial modes for $M=1.0\,\mathrm{M}_\odot$ and $1.8\,\mathrm{M}_\odot$ evolutionary models are shown in Figure \ref{fig2} as filled and open symbols, respectively. Circles, triangles, inverse triangles, squares, and diamonds represent the pulsationally unstable radial fundamental modes, the first, second, third, and fourth overtones, respectively. In our computation of non-adiabatic radial oscillations, almost all of the p1--p39 modes of all the RGB and AGB models converged successfully without difficulty. Only for high-luminosity AGB models, the non-adiabatic oscillations of the fundamental modes were often nonconvergent, leading to the missing modes in Figure \ref{fig2}. We are still not very clear whether the non-convergence of the fundamental modes came from numerical calculations or had other causes. If the reason is the former, it it very difficult to explain why the non-convergence only existed in the calculations of the fundamental modes, while the other modes converged exceptionally well.

As can be seen in Figure \ref{fig2}, the theoretical period ranges of low-order radial oscillations are approximately consistent with the observed period ranges of the LPV sequences. However, there are systematic differences between the observed and theoretical PL relations. The main reasons are as follows.

\begin{enumerate}
\item Uncertainties in the transformation from theoretical stellar parameters ($M$, $L$, $T_\mathrm{e}$) to observed magnitudes ($V$, $I$, $K_\mathrm{S}$).

The uncertainties are more significant for luminous RGB and AGB stars, which have extended dust envelopes. By comparing Figures~\ref{fig2}a \& b, it is can be seen that the differences between theoretical and observed PL relations in $K_\mathrm{S}$ are clearly larger than those in $W_I$. It seems that the long-wavelength end is more affected by the dust envelopes.

\item Uncertainties in stellar evolutionary models.

Compared with normal-mass stars in their early evolutionary stages, the evolutions of RGB and AGB stars are subject to uncertainties from mass loss, convection treatment, and low-temperature opacity.

\item Mass variations.

The luminous red variables observed by MACHO and OGLE are a group of stars with certain mass, age, and metallicity distributions, while our theoretical stellar models shown in Figure~\ref{fig2} are RGB and AGB models with $M=1.0\,\mathrm{M}_\odot$ and $Z=0.008$. Therefore differences between theory and observation are expected. Using a modified Petersen diagram, \citet{Wood2015} found that the stars on sequences A$'$--C have different masses.
\end{enumerate}

\section{Excitation mechanism}\label{sec4}

We have long known that Cepeheid and Cepeheid-like variables are driven by radiative $\kappa$ mechanism, and the existence of the red edge of the Cepheid instability strip is due to convective damping. Therefor for a long time convection is believed to be a pure damping mechanism of oscillations. However, why are there still various types of red variables in the low-temperature area beyond the Cepheid instability strip? \citet{XDC1998} studied non-adiabatic oscillations of luminous red variables, and pointed out the existence of a Mira instability strip in the low-temperature area beyond the Cepheid instability strip.

In low-temperature red variables, convection replaces radiation as the dominant means of energy transport. Radiation is still an important driving and damping mechanism of oscillations in red variables, but it works together with convection, instead of being the dominant mechanism in high-temperature variables. Radiative transfer is a well-know physical process, while the coupling between convection and oscillations remains a major difficulty in the study of red variables \citep{HD2015}. Therefore it is fair to say that convection is the key to solve the problem of the excitation of oscillations in red variables. Unfortunately, so far there has not been a widely accepted stellar convection theory. Therefore, controversies are unavoidable about the excitation mechanism of oscillations in red variables. The following discussions represent the theoretical interpretations of this problem within the framework of our non-local and time-dependent convection theory \citep{tcpss1}

Accumulated work is a convenient and useful way in studying the excitation mechanisms of pulsating variables. It not only gives quantitatively estimate of the plusational amplitude growth rate, but also shows the locations of the driving and damping. The total accumulated work can be written as
\begin{equation}
W=W_{P_\mathrm{g}}+W_{P_\mathrm{t}}+W_\mathrm{vis},
\label{eq:w}
\end{equation}
where $W_{P_\mathrm{g}}$, $W_{P_\mathrm{t}}$, and $W_\mathrm{vis}$ are the gas pressure component, the turbulent pressure component, and the turbulent viscosity component, respectively.
\begin{equation}
W_{P_\mathrm{g}}=-\frac{\pi}{2E_\mathrm{k}}\int_0^{M_0}\mathrm{I}_\mathrm{m}\left\{\frac{P}{\rho}\frac{\delta P}{P}\frac{\delta\rho^*}{\rho}\right\}\mathrm{d}M_r,
\label{eq:wpg}
\end{equation}
\begin{equation}
W_{P_\mathrm{t}}=-\frac{\pi}{2E_\mathrm{k}}\int_0^{M_0}\mathrm{I}_\mathrm{m}\left\{\rho x^2\frac{\delta x}{x}\frac{\delta\rho^*}{\rho}\right\}\mathrm{d}M_r,
\label{eq:wpt}
\end{equation}
\begin{equation}
W_\mathrm{vis}=-\frac{\pi}{2E_\mathrm{k}}\int_0^{M_0}\mathrm{I}_\mathrm{m}\left\{\rho \chi^{1}_{1}\frac{\delta\chi_{1}^{1*}}{\chi_1^1}\frac{\mathrm{d}}{\mathrm{d}\ln r}\left(\frac{\delta r^*}{r}\right)\right\}\mathrm{d}M_r,
\label{eq:wvis}
\end{equation}
where
\begin{equation}
E_\mathrm{k}=\frac{1}{2}\int_0^{M_0}\rho\omega^2\delta r\delta r^*\mathrm{d}M_r,
\label{eq:ek}
\end{equation}
and $\omega$ is the circular frequency of the oscillation mode. $\rho g^{ij}x^2$ and $\rho\chi^{ij}$ are the isotropic and anisotropic components of the turbulent Reynolds stress and are expressed in tensor form with $g^{ij}$ being the metric tensor (see Paper I and \citet{ZDX2017} for related definitions and discussions).

Convection drives mixing of material and transport and exchange of energy and momentum in stellar interiors, and therefore affects the structure, evolution, and pulsation stability of stars. Equations \ref{eq:wpt} and \ref{eq:wvis} are the work contributions by the Reynolds stress by means of momentum exchange, i.e. the dynamic coupling between convection and oscillations. Due to the inertia of turbulent motions, the variation of the turbulent pressure $\rho x^2$ in stellar oscillations always lags slightly behind the variation of density. As a result, a positive Carnot cycle is formed in the pressure-volume ($P_\mathrm{t}-V$) diagram, converting irregular turbulent kinetic energy into regular kinetic energy of stellar oscillations. Therefore, turbulent pressure is always a driving mechanism of oscillations. It has important effects on the excitation of stellar oscillations \citep{XD2013, tcpss2, Antoci2014}.

The anisotropic component $\chi^{ij}$ represents shear motions of the fluid, converting regular kinetic energy of oscillations into irregular turbulent kinetic energy. This process takes place at low wave numbers (large-scale turbulence) of the turbulent energy spectrum. The energy is then gradually shifted to high wave numbers (small-scale turbulence) as a result of turbulence cascade, and is eventually converted into thermal energy by molecular viscosity. Therefore, turbulent viscosity is always a damping mechanism against oscillations.

The gas pressure component of the accumulated work in equation \ref{eq:wpg} includes the contributions from both the radiative and convective energy transfer. We can separate the contributions from the radiative flux, the turbulent thermal flux, and the turbulent kinetic energy flux one by one in the average energy equation \citep{XD2010, XD2013}. They are closely coupled to each other; variation in one causes compensating variations in the others. In the deep interior of the convection zone, the convective flux dominates. Our study shows that the variation of the turbulent thermal flux is usually synchronized with the variation of density (with a slight phase lag). This means that in stellar oscillations thermal convection takes out more energy at high-temperature high-density phase, and blocks in more energy at low-temperature low-density phase. This mechanism works like a refrigerating machine, converting pulsation kinetic energy into thermal energy. Therefore, turbulent thermal convection is usually a damping mechanism against oscillations.

\begin{figure}
\includegraphics[width=\columnwidth]{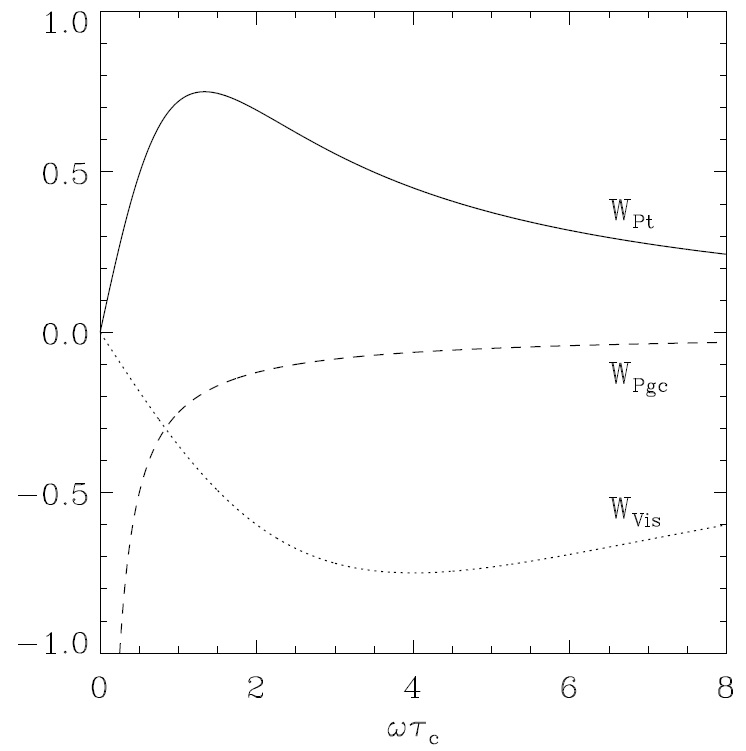}
\caption{Sketch of the frequency dependence of the effects of turbulent pressure (solid line), turbulent thermal flux (dashed line), and turbulent viscosity (dotted line).}
\label{fig3}
\end{figure}

\begin{figure}
\includegraphics[width=\columnwidth]{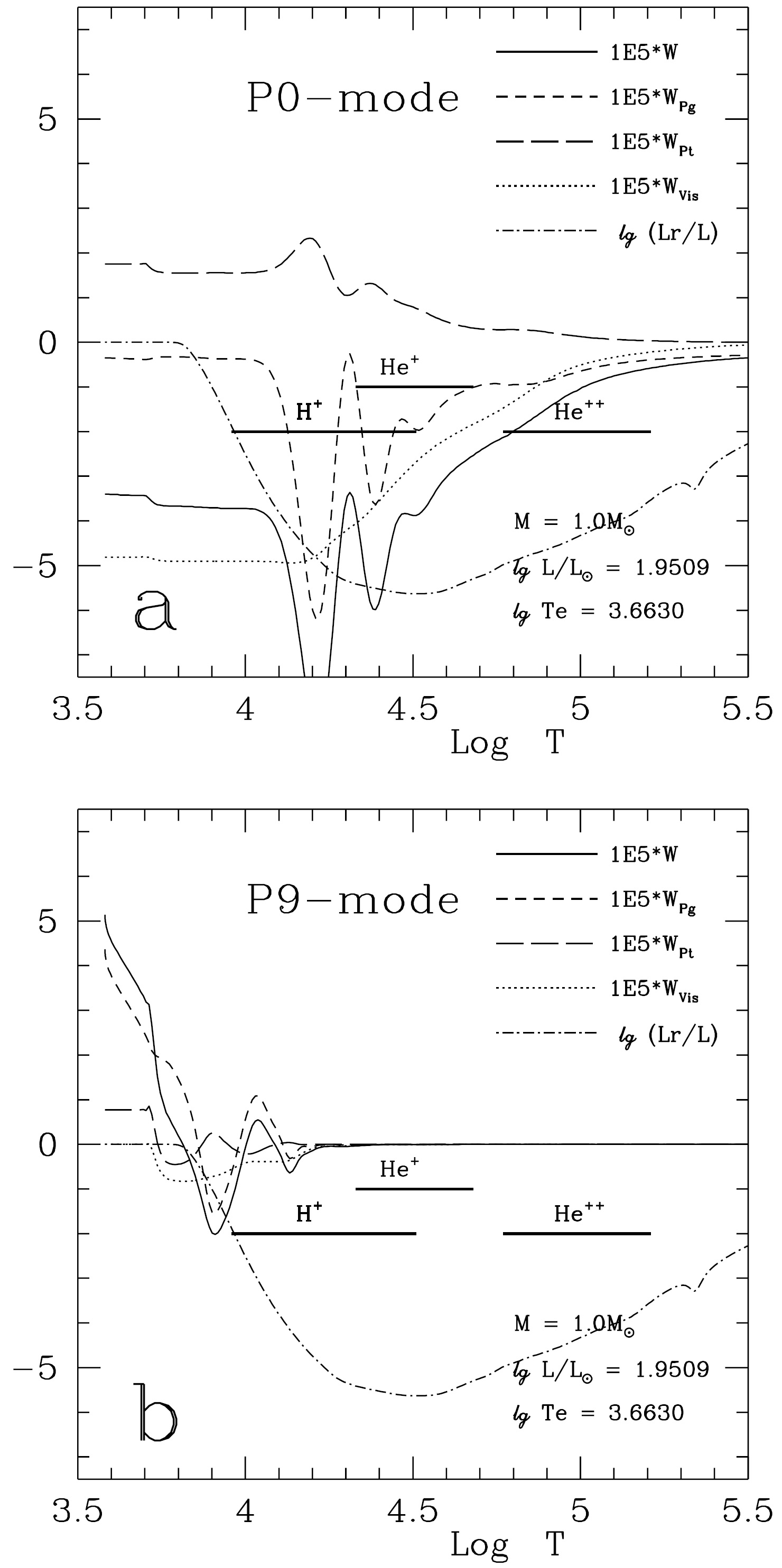}
\caption{Accumulated work and its components as a function of depth $\log T$ for a low-luminosity red giant. The solid lines are the total accumulated work $W$. The dashed, long-dashed, and dotted lines are the gas component $W_{P_\mathrm{g}}$, turbulent pressure component $W_{P_\mathrm{t}}$, and turbulent viscosity component $W_\mathrm{vis}$, respectively. The dash-dotted lines are the fractional radiative flux $\log L_\mathrm{r}/L$. The horizontal lines indicate the locations of the ionization regions of hydrogen and helium. Panel a: p0 mode. Panel b: p9 mode.}
\label{fig4}
\end{figure}

\begin{figure}
\includegraphics[width=\columnwidth]{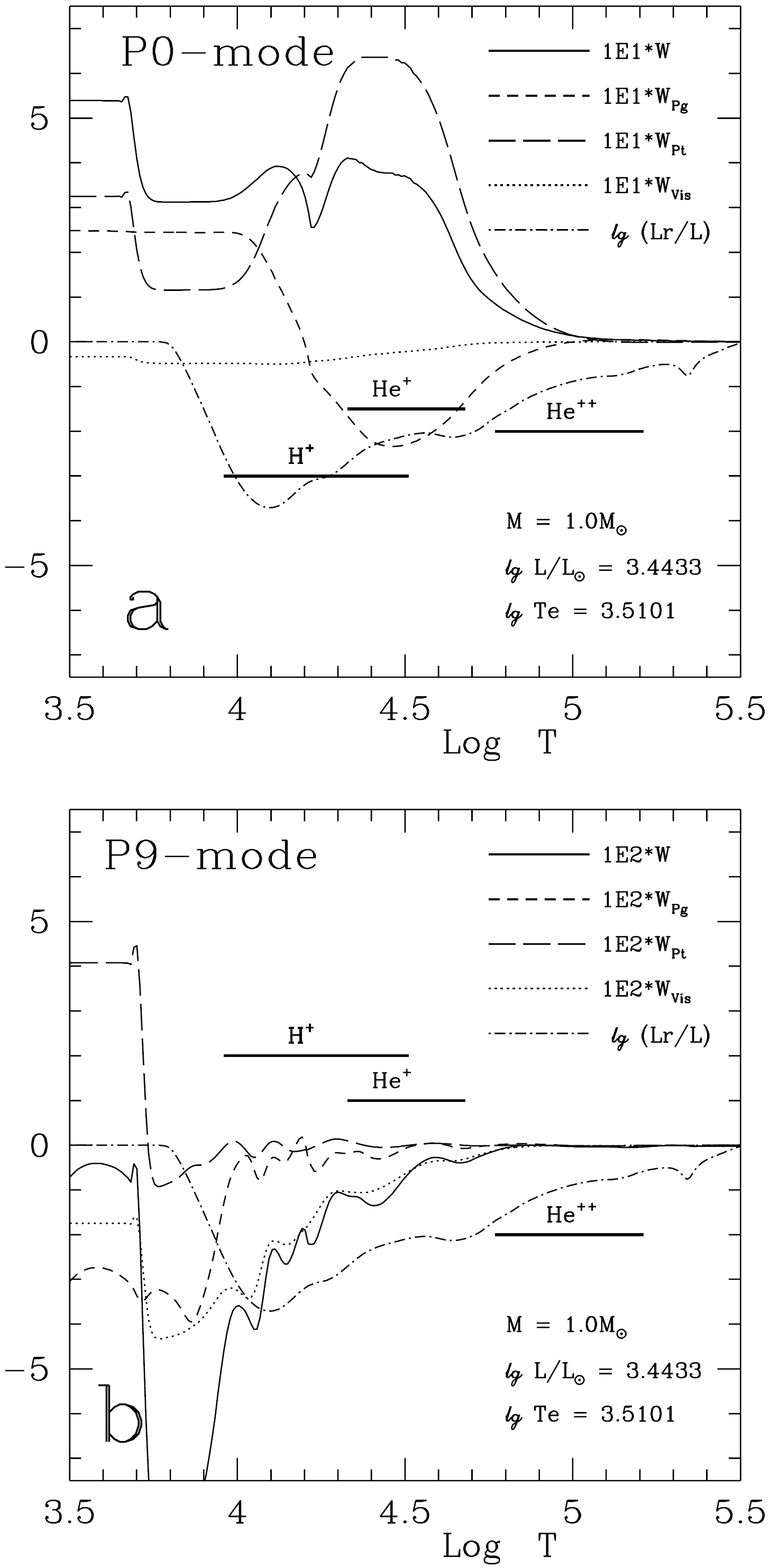}
\caption{The same as Figure \ref{fig3} but for a high-luminosity red giant.}
\label{fig5}
\end{figure}

Figure \ref{fig3} is a sketch showing the turbulent pressure component $W_{P_\mathrm{t}}$, the turbulent viscosity component $W_\mathrm{vis}$, and the turbulent thermal convection component $W_{P_\mathrm{gc}}$ as a function of $\omega\tau_\mathrm{c}$, where $\tau_\mathrm{c}$ is the dynamic time scale of convective motions \citep{XD2013}. As stellar parameters ($M$, $L$, $T_\mathrm{e}$, $X$, $Y$, $Z$) and oscillation frequencies vary, the relative sizes of the contributions from radiation, turbulent pressure, turbulent thermal convection, and turbulent viscosity also vary. The combined effect is that the total accumulated work $W$ sometimes is positive, and sometimes is negative, corresponding to unstable and stable oscillations, respectively. Therefore a star shows different oscillation characteristics in different evolutionary stages. Figure \ref{fig4} and \ref{fig5} show the accumulated work $W$ (solid lines) and the components $W_{P_\mathrm{g}}$ (dashed lines) $W_{P_\mathrm{t}}$ (long-dashed lines) $W_\mathrm{vis}$ (dotted lines) of the fundamental modes (panels a) and the ninth overtones (panels b) as a function of depth for a low-luminosity red giant and a high-luminosity red giant, respectively. The fractional radiative flux $\log L_\mathrm{r}/L$ (dashed-dotted lines) and the hydrogen and helium ionization zones (thick horizontal lines) are also plotted. In the low-luminosity red giant, the fundamental mode (Figure \ref{fig4}a) is stable due to the damping mainly from turbulent viscosity, but the ninth overtone (Figure \ref{fig4}b) is unstable with the driving contributed by both the gas pressure and turbulent pressure components. However, we see the opposite results in the high-luminosity red giant. The fundamental mode (Figure \ref{fig5}a) is pulsationally unstable due to the dominating driving effect from turbulent pressure; while the ninth overtone (Figure \ref{fig5}b) becomes stable because the damping effects from both turbulent thermal convection and turbulent viscosity overtake the driving effect from turbulent pressure.

\begin{figure}
\includegraphics[width=\columnwidth]{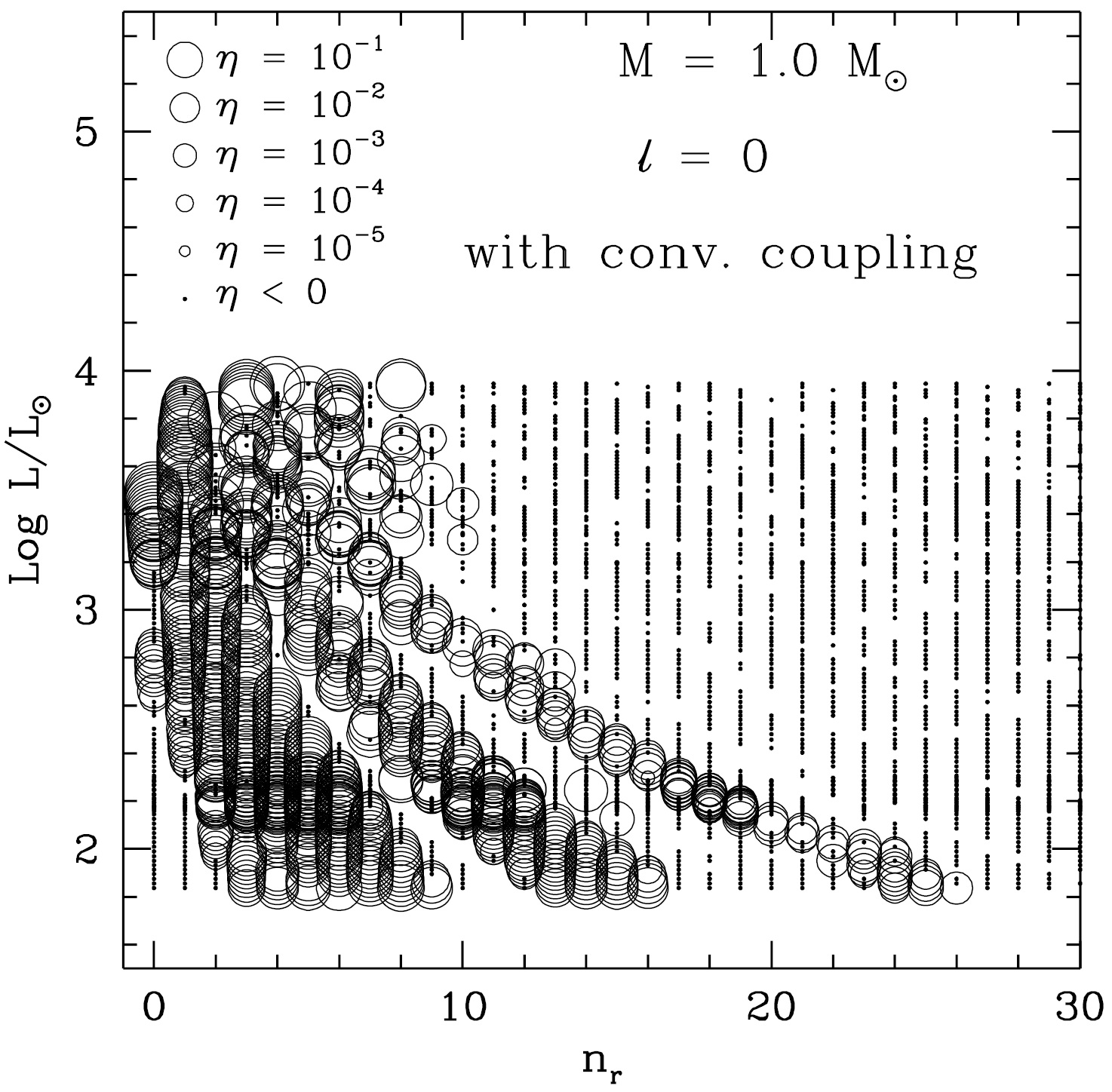}
\caption{Pulsationally stable (small dots) and unstable (open circles) radial modes in the $\log L/\mathrm{L}_\odot-n_\mathrm{r}$ plane for AGB models. The sizes of the open circles are proportional to the logarithms of the amplitude growth rates of the modes.}
\label{fig6}
\end{figure}

Figure \ref{fig6} shows the numerical results of linear non-adiabatic oscillations for $1.0\mathrm{M}_\odot$ stellar models. The small dots are pulsationally stable modes, and the open circles are unstable modes. The sizes of the open circles are proportional to the logarithms of the amplitude growth rates. It can be clearly seen that low-order modes are pulsationally stable in low-luminosity red giants, while the mid- to high-order modes are pulsationally unstable. Therefore low-luminosity red giants show characteristics of solar-like oscillations. As luminosity increases, pulsationally unstable modes shift to lower radial orders. In high-luminosity red giants, only a few low-order modes are pulsationally unstable, while all the mid- to high-order modes become pulsationally stable. Therefore high-luminosity red giants show typical characteristics of Mira-like oscillations.

\section{Summary}\label{sec5}

We have computed the linear non-adiabatic oscillations of evolutionary models of low-mass red giants. The main results can be summarized as follows.
\begin{enumerate}
\item Sequences A--C of luminous red variables in LMC of MACHO and OGLE observations are low-order radial modes of low-mass RGB and AGB stars.

\item Oscillations in Miras and OSARGs can be self-excited as a result of radiation and the coupling between convection and pulsation. Turbulent pressure has important effects on the excitation of oscillations in low-temperature red variables.

\item Our study shows that for low-luminosity red giants, the low-order modes are stable, while the intermediate- and high-order modes are unstable. These stars show characteristics of solar-like oscillations. As the luminosity increases, unstable modes move towards lower orders. In luminous red giants, only a few low-order radial modes are unstable, while all of the intermediate- and high-order modes become stable. These stars show characteristics of Mira-like oscillations.
\end{enumerate}

\section*{Acknowledgements}
This work was supported by National Natural Science Foundation of China (NSFC) through grants 11373069, 11473037, and 11403039.




\begin{thebibliography}{99}
\bibitem[\protect\citeauthoryear{Alexander \& Ferguson}{1994}]{LTO94}Alexander D. R., Ferguson J. W., 1994, \apj, 437, 879

\bibitem[\protect\citeauthoryear{Antoci et al.}{2014}]{Antoci2014}Antoci V., Cunha M., Houdek G., et al., 2014, \apj, 796, 118

\bibitem[\protect\citeauthoryear{Bertelli et al.}{2008}]{Padova2008}Bertelli G., Girardi L., Marigo P., Nasi E., 2008, \aap, 484, 815

\bibitem[\protect\citeauthoryear{Bertelli et al.}{2009}]{Padova2009}Bertelli G., Nasi E., Girardi L., Marigo P., 2009, \aap, 508, 355

\bibitem[\protect\citeauthoryear{B\"{o}hm-Vitense}{1958}]{MLT} B{\"o}hm-Vitense, E. 1958, \zap, 46, 108

\bibitem[\protect\citeauthoryear{Carpenter}{2001}]{Carpenter2001}Carpenter J. M., 2001, \aj, 121, 2851

\bibitem[\protect\citeauthoryear{Christensen-Dalsgaard et al.}{2001}]{CDKM2001}Christensen-Dalsgaard J., Kjeldsen H., Mattei J. A., 2001, \apjl, 562, L141

\bibitem[\protect\citeauthoryear{D\"appen et al.}{1988}]{MHD3}D\"appen W., Mihalas D., Hummer D. G., Mihalas B. W., 1988, \apj, 332, 261

\bibitem[\protect\citeauthoryear{Deng et al.}{2006}]{DXC2006}Deng L., Xiong D. R., Chan K. L., 2006, \apj, 643, 426

\bibitem[\protect\citeauthoryear{Henyey et al.}{1964}]{HFG1964}Henyey L. G., Forbes J. E., Gould N. L., 1964, \apj, 139, 306

\bibitem[\protect\citeauthoryear{Houdek \& Dupret}{2015}]{HD2015}Houdek G., Dupret M.-A., 2015, Living Rev. Solar Phys., 12, 8

\bibitem[\protect\citeauthoryear{Hummer \& Mihalas}{1988}]{MHD1}Hummer D. G., Mihalas D., 1988, \apj, 331, 794

\bibitem[\protect\citeauthoryear{Ita et al.}{2004}]{ITM2004}Ita Y., Tanab\'e T., Matsunaga N., et al., 2004, \mnras, 347, 720

\bibitem[\protect\citeauthoryear{Kholopov et al.}{1985--1988}]{GCVS}Kholopov P. N., Samus N. N., Frolov M. S., et al., 1985--1988, General Catalogue of Variable Stars 4th edn. Nauka, Moscow

\bibitem[\protect\citeauthoryear{Kiss \& Bedding}{2003}]{KB2003}Kiss L. L., Bedding T. R., 2003, \mnras, 343, L79

\bibitem[\protect\citeauthoryear{Mihalas et al.}{1988}]{MHD2}Mihalas D., D\"appen W., Hummer D. G., 1988, \apj, 331, 815

\bibitem[\protect\citeauthoryear{Pietrzy\'nski et al.}{2013}]{PGG2013}Pietrzy\'nski G., Graczyk D., Gieren W., et al., 2013, \nat, 495, 76
\bibitem[\protect\citeauthoryear{Rogers \& Iglesias}{1992}]{OPAL92}Rogers F. J., Iglesias C. A., 1992, ApJS, 79, 507

\bibitem[\protect\citeauthoryear{Soszy\'nski et al.}{2007}]{SDU2007}Soszy\'nski I., Dziembowski W. A., Udalski A., et al., 2007, \actaa, 57, 201

\bibitem[\protect\citeauthoryear{Soszy\'nski et al.}{2004}]{SUK2004}Soszy\'nski I., Udalski A., Kubiak M., et al., 2004, \actaa, 54, 129

\bibitem[\protect\citeauthoryear{Soszy\'nski et al.}{2005}]{SUK2005}Soszy\'nski I., Udalski A., Kubiak M., et al., 2005, \actaa, 55, 331

\bibitem[\protect\citeauthoryear{Soszy\'nski et al.}{2009}]{SUS2009}Soszy\'nski I., Udalski A., Szyma\'nski M. K., et al., 2009, \actaa, 59, 239

\bibitem[\protect\citeauthoryear{Takayama et al.}{2013}]{TSI2013}Takayama M., Saio H., Ita Y., 2013, \mnras, 431, 3189

\bibitem[\protect\citeauthoryear{Trabucchi et al.}{2017}]{Trabucchi2017}Trabucchi M., Wood, P. R., Montalb\'an J., et al., 2017, \apj, 847, 139

\bibitem[\protect\citeauthoryear{Wood}{2000}]{Wood2000}Wood P. R., 2000, \pasa, 17, 18

\bibitem[\protect\citeauthoryear{Wood}{2015}]{Wood2015}Wood P. R., 2015, \mnras, 448, 3829

\bibitem[\protect\citeauthoryear{Wood et al.}{1999}]{Wood1999}Wood P. R., Alcock C., Allsman R. A., et al., 1999, in Le Betre T., Lebre A., Waelkens C., eds, Proc. IAU Symp. 191, Asymptotic Giant Branch Stars. Kluwer, Dordrecht, p. 151

\bibitem[\protect\citeauthoryear{Worthey \& Lee}{2011}]{WL2011}Worthey G., Lee H.-C., 2011, \apjs, 193, 1

\bibitem[\protect\citeauthoryear{Wray et al.}{2004}]{WEP2004}Wray J. J., Eyer L., Paczy\'nski B., 2004, \mnras, 349, 1059

\bibitem[\protect\citeauthoryear{Xiong}{1989}]{X1989}Xiong D. R., 1989, \aap, 209, 126

\bibitem[\protect\citeauthoryear{Xiong \& Deng}{2010}]{XD2010}Xiong D. R., Deng L., 2010, \mnras, 405, 2759

\bibitem[\protect\citeauthoryear{Xiong \& Deng}{2013}]{XD2013}Xiong D. R., Deng L., 2013, Res. in Astron. and Astrophys., 13, 1269

\bibitem[\protect\citeauthoryear{Xiong et al.}{1997}]{XCD1997}Xiong D. R., Cheng Q. L., Deng L., 1997, \apjs, 108, 529

\bibitem[\protect\citeauthoryear{Xiong et al.}{1998}]{XDC1998}Xiong D. R., Deng L., Cheng Q. L., 1998, \apj, 499, 355

\bibitem[\protect\citeauthoryear{Xiong et al.}{2015}]{tcpss1}Xiong D. R., Deng L., Zhang C., 2015, \mnras, 451, 3354 (Paper I)

\bibitem[\protect\citeauthoryear{Xiong et al.}{2016}]{tcpss2}Xiong D. R., Deng L., Zhang C., Wang K., 2016, \mnras, 457, 3163

\bibitem[\protect\citeauthoryear{Zhang et al.}{2017}]{ZDX2017}Zhang C., Deng L., Xiong D. R., RAA, 17, 29

\end{thebibliography}



\bsp	
\label{lastpage}
\end{document}